\documentclass[10pt,a4paper]{article}
\usepackage{amssymb}
\usepackage{graphicx}
\usepackage{layout}
\usepackage{amssymb}
\begin{document}

\author{Matteo Luca Ruggiero$^{\S,\ast}$
\\ \\
\small
$^\S$ Dipartimento di Fisica, Politecnico di Torino,\\
\small Corso Duca degli Abruzzi 24, 10129 Torino, Italy \\
\small $^\ast$ INFN, Via P. Giuria 1, 10125 Torino, Italy\\ \\
 E-mail: matteo.ruggiero@polito.it}
\title{The Relative Space: Space  Measurements on a Rotating Platform}
\maketitle

\normalsize
\begin{abstract}
We introduce here the concept of relative space, an extended
3-space which is recognized as the only space having an
operational meaning in the study of the space geometry of a
rotating disk. Accordingly, we illustrate how space measurements
are performed in the relative space, and we show that an old-aged
puzzling problem, that is the Ehrenfest's paradox, is explained in
this purely relativistic context. Furthermore, we   illustrate the
kinematical origin of the tangential dilation which is responsible
for the solution  of the Ehrenfest's paradox.
\end{abstract}

\section{Introduction}\label{sec:intro}

In the Special Theory of Relativity (SRT) the rotation of the
reference frame, contrary to the translation, has an absolute
character and can be locally measured by the Foucault's pendulum
or by the Sagnac experiment. Indeed, this peculiarity of rotation,
inherited by Newtonian physics, is difficult to understand in a
relativistic context. As a matter of fact, many authors who were
contrary to SRT, had found, in the relativistic approach to
rotation, important arguments against the self-consistency of the
theory. Already in 1909 Ehrenfest\cite{ehrenfest} pointed out an
internal contradiction in  SRT, applied to the case of a rotating
disk; few years later, in 1913, Sagnac\cite{sagnac},\cite{post}
evidenced an apparent contradiction in SRT with respect to the
experimental data.

Since those years, these seminal papers had influenced discussions
on the foundations of SRT, even if these "paradoxes" disappear
when a careful analysis is undertaken, following the very axioms
of the theory. However, it may appear surprising that, even after
one century, in some papers or in textbooks, misleading or even
uncorrect arguments are given to explain the Ehrenfest's paradox
and the Sagnac effect. For instance, Klauber in 1998
\cite{klauber},\cite{klauber2} proposed a ''New Theory of Rotating
Frames'', in order to amend the contradictions which, according to
him, appear in SRT when applied to rotating frames.

Elsewhere\cite{rizzi02} we carefully studied  the Ehrenfest's
paradox, showing that it can be solved on the bases of purely
kinematical arguments in SRT. To this end, we adopted a
geometrical approach, based on (i) a precise definition of the
concept of ''space of the disk'', (ii) a precise choice of the
''standard rods'' used by the observers on the platform.  The
space of the disk has been formally identified with what we called
the "relative space"; its geometry has been recognized to be
non-Euclidean\footnote{The non null curvature of the space of the
disk has nothing to do with the spacetime curvature, which is
always null  if gravitation is not taken into account.} and its
metric coincides with the one which is found in classic textbooks
of relativity, in spite of a shift of the context, that we have
stressed. We adopted the projection technique introduced by
Cattaneo\cite{cattaneo}, \cite{catt1}, \cite{catt2}, \cite{catt3},
\cite{catt4} to approach the problem, since  this allows an
elegant and straightforward description of the  geometry of any
reference frame.

Even though a global  \textit{isotropic} $1+3$ splitting  of
space-time is not possible when we deal with rotating observers,
the introduction of the relative space allows  well defined
procedures for space  measurements that can be actually performed
globally by the observers in  rotating frames and which reduce to
the standard measurements in any locally co-moving inertial
frames.

Here we are going to show that these procedures allow a systematic
study of  the space measurements, which makes the Ehrenfest's
paradox disappear. Moreover, we are going to illustrate in a clear
way (with the aid of a pictorial representation) the kinematical
origin of the dilation which leads to the solution of the
Ehrenfest's paradox.

According to us the concept of relative space can help to make rid
of lots of misunderstandings, often caused by the lack of proper
definitions of some crucial concepts in the theoretical apparatus
used to describe these physical problems: indeed, they  may appear
puzzling or contradictory in SRT only when ambiguous entities and
procedures are adopted.

\section{The Ehrenfest's Paradox}\label{sec:paraE}

According to Ehrenfest\cite{ehrenfest}, the formulation of the
paradox is the following one:

Let $R,R'$ be the radii of the  rotating disk, as measured,
respectively, by the inertial and rotating observer; $\omega$ is
the constant angular velocity of the disk, as measured in the
inertial frame. The paradox arises when the following
contradictory statements are taken into account:

\indent (a) The circumference of the disk must show a contraction
relative to its rest state, $2 \pi R < 2 \pi R^{\prime}$, since
each element
of the circumference moves in its own direction with instantaneous speed $%
\omega R$. \newline
\newline
\indent (b) If one considers an element of a radius, its
instantaneous velocity is perpendicular to its length; thus, an
element of the radius
cannot show a contraction with respect to the rest state. Therefore $%
R=R^{\prime}$
\newline

On the bases of this contradiction, Ehrenfest pointed out the
apparent inconsistence of the kinematics of bodies which are
rigid, according to the definition of rigidity given by Born (see
below).\\

Despite the great number of authors who tried to explain the
paradox, the most popular attempts of solution can be summarized
in this way:

\begin{itemize}
\item[(1)] neither the rods along the rim nor the circumference do
contract; neither the rods along the radius nor the radius itself
do contract; as a consequence, the space of the rotating disk is
Euclidean;\\
\textit{(See f.i. Tartaglia\cite{tartaglia},
Klauber\cite{klauber},\cite{klauber2})}

\item[(2)] both the rods along the rim and the circumference
contract; neither the rods along the radius nor the radius itself
do contract; as a consequence the surface of the disk bends,
because of rotation.\\
\textit{(Sted-Donaldson\cite{stedon}, Galli\cite{galli})}

\item[(3)] rods along the rim do contract, while the circumference
does not; neither the rods along the radius nor the radius itself
do contract. As a consequence the space of the disk is not
Euclidean: $\Longrightarrow 2\pi R=L<L^{\prime }=\gamma 2\pi
R^{\prime }$,
where $\gamma $ is the Lorentz factor;\\
\textit{(Einstein\cite{einstein}, Berenda\cite{berenda},
Arzeli\`{e}s\cite{arzelies}, Gr\o n\cite{gron3},\cite{gron1}, M\o
ller \cite{moller}, Landau-Lifshitz\cite{landau1})}
\end{itemize}

Let us say few words about why these solution are not completely
satisfactory\footnote{See \cite{rizzi02} for a historical and
detailed analysis of these attempts of solutions of the
paradox.}.\\

The approach (1) is, indeed, the one which is in agreement with
the common sense, since people can hardly figure out how
non-Euclidean features may appear on a disk because of its
rotation. However, to maintain the statement "nothing happens on a
rotating disk" a remarkable price must be paid. For instance,
Klauber analysis is based on a deep criticism of the foundations
of SRT: he claims that SRT cannot be applied to rotating frames,
but it must be deeply amended to take into account the non
inertial motion of rotating observers. In particular, he maintains
that the "Hypothesis of Locality"(see Mashhoon\cite{mashh}), which
states the local equivalence of an accelerated observer with a
momentarily co-moving inertial observer\footnote{Provided that
standard rods and clocks are used.}, is not valid in rotating
frames. Accordingly, his approach appears a challenge to the very
foundations of relativity, because the Hypothesis of Locality is
one of the most important axioms of theory. Dropping this axiom
(which ultimately is justified by the experimental observations)
forces us to abandon the idea that the theory of relativity can
describe the physical world, since, actually, there are no
perfectly inertial frames in the real world\footnote{As pointed
out by Selleri\cite{sellerilibro}, "because of the terrestrial
rotation, the orbital motion of the Earth around the Sun, the
Galactic rotation... all of our knowledge about inertial systems
have been obtained in frames having a small but non zero
[centripetal] acceleration".}.

On the other hands, Tartaglia's result can be justified and
accepted if and only if a specific choice of the measuring rods is
done. However, this choice is questionable when applied to the
case of a rotating disk: Tartaglia's measuring rods are not the
standard rods of SRT. As we shall show in Sec. \ref{sec:georot}
below, in SRT we can locally substitute the light rays for the
standard rigid rods; furthermore,  we shall show in Sec.
\ref{sec:lengths} that a careful  study of the effects of an
acceleration process  explains the kinematical origin of the
Ehrenfest paradox.

The approach (2) introduces difficulties in the explanation of the
paradox (for instance  the dynamical properties of the rotating
disk must be taken into account) and inconsistencies:  a non
symmetric deformation with respect to the plane of the
non-rotating disk should determine a screw sense in space, thus
violating spatial parity \textit{in a purely kinematical context}.

Finally, the approach (3) formally agrees with ours: however,
those authors do not define explicitly the geometric context in
which measurements are made. Moreover, some of them refer to the
"space of the disk" as if it were a submanifold or a subspace
embedded in space-time: this is not the case, since the lack of
time-orthogonality (see below) does not allow a global $1+3$
splitting. Here, we are going to show that the relative space is
the only extended space, having a  clear operational meaning, that
can be formally defined as the physical space of the disk.

\section{The Space-Time geometry of a rotating platform} \label{sec:georot}

\subsection{Parameterizing the rotating frame}\label{ssec:param}

The physical space-time is a (pseudo)riemannian manifold $%
\mathcal{M}^{4}$, that is a pair $\left(
\mathcal{M},\mathbf{g}\right)$, where $\mathcal{M}$ is a connected 4-dimensional Haussdorf manifold and $%
\mathbf{g}$ is the metric tensor\footnote{%
The riemannian structure implies that $\mathcal{M}$ is endowed
with an affine connection compatible with the metric, i.e. the
standard Levi-Civita connection.}. Let the signature of the
manifold be $(1,-1,-1,-1)$. Suitably differentiability condition,
on both $\mathcal{M}$ and $\mathbf{g}$, are assumed.

Let  $K$ be the inertial laboratory frame, in which the platform
(with  its measuring apparatus) rotates with a constant angular
velocity $\omega$. Let $K$ be parameterized by a set of
coordinates $\left\{ x^{\mu }\right\}~=~\left(ct,r,\vartheta
,z\right)$, where $t$ is the inertial time of $K$ and $\left(
r,\vartheta,z\right)$ are the cylindrical spatial coordinates.

In this frame, let us consider the equations
\begin{equation}
\left\{
\begin{array}{rcl}
r & = & r_{0} \\
\vartheta  & = & \vartheta _{0}+\omega t \\
z & = & z_{0}
\end{array}
\right. \mathrm{\ .}  \label{discorotante}
\end{equation}
If $r_{0}\in [0,R]\,,$ these equations describe the points of a
cylinder with radius $R$, rotating with constant angular velocity
$\omega $. When $z_{0}$ is the same for each point of the system,
we deal with a rotating disk, whose points have cylindrical
coordinates $(r_0,\vartheta_0,z_0)$, at the initial time
$t=0$.\\

The world-lines of each point of the disk are time-like helixes
(whose pitch, depending  on $\omega $, is constant), wrapping
around the cylindrical surface  $r~=~r_{0}~=~const$, with $r \in
[0,R]$. These helixes fill, without intersecting, the whole
space-time region defined by $r\leq R$;\footnote{The condition
$R<c/\omega $ is usually imposed: this simply means that the
velocity of the points of the disk cannot reach the velocity of
light.} they constitute a time-like "congruence"
$\Gamma$ which defines the rotating frame $K_{rot}$%
, at rest with respect to the disk\footnote{The concept of
"congruence" refers to a set of word-lines filling the manifold,
or some part of it, smoothly, continuously and without
intersecting.}.\\

Let us introduce the coordinate transformation
\begin{equation}
\left\{
\begin{array}{rcl}
x'^{0} & = & ct^{\prime }=ct \\
x'^{1} & = & r^{\prime }=r \\
x'^{2} & = & \vartheta ^{\prime }=\vartheta -\omega t \\
x'^{3} & = & z^{\prime }=z
\end{array}
\right. \mathrm{\ .}  \label{catt}
\end{equation}

This coordinate transformation  defines the passage from a chart
$\left\{ x^{\mu}\right\}$ adapted to the inertial frame $K$ to a
chart  $\left\{ x'^{\mu}\right\}$ adapted to the rotating frame
$K_{rot}$\footnote{If $\{x^{\mu }\}=(x^{{0}},x^{1},x^{2},x^{3})$
is a system of coordinates in a subset $U \subset \mathcal{M}$,
these coordinates are said to be "adapted" to the physical frame
if
\begin{displaymath}
g_{{00}}>0\ \ \ \ g_{ij }dx^{i }dx^{j }<0 \label{eq:admiss}
\end{displaymath}
(Greek indices run from 0 to 3, Latin indices run from 1 to 3).
Accordingly, the couple $\left(\{x^\mu\},U\right)$ is said to be a
"chart" adapted to the physical frame. The subset $U$ is the
coordinate domain of this chart; in our case, $U$ is defined by
the condition $r\in ]0, R]$. In what follows, we shall always
refer to this domain, even if it will not be explicitly declared.}
.

In the chart $\left\{ x'^{\mu}\right\}$ the metric tensor is
written in the form\cite{rizzi02}:
\begin{equation}
g_{\mu \nu }^{\prime }=\left(
\begin{array}{cccc}
1-\frac{\omega ^{2}{r'}^{2}}{c^{2}} & 0 & -\frac{\omega {r'}^{2}}{c} & 0 \\
0 & -1 & 0 & 0 \\
-\frac{\omega {r'}^{2}}{c} & 0 & -{r'}^{2} & 0 \\
0 & 0 & 0 & -1
\end{array}
\right)  \label{born}
\end{equation}
This is the so called "Born metric".

In the chart $\left\{ x'^{\mu \ }\right\} $ the time  $t'$ is
equal to the coordinate time $t$ of the inertial frame $K$. In
this way, we label  each event $P$ in $K_{rot}$ using the time of
a clock at rest in $K$, whose world-line (a straight line parallel
to the time axis) intersects $P$, and not by means of a clock at
rest on the disk. The transformation (\ref{catt}) has a Galilean
character, and this is due to the peculiarity of angular velocity
which, contrary to translational velocity, has an absolute value,
that can be locally measured.\\

\textbf{Remark  }The parameterization of the rotating frame
$K_{rot}$ by the coordinate $t$ of the inertial frame $K$ is the
only way to synchronize the clocks (globally) on the platform: in
fact their proper times cannot be synchronized by  Einstein's
convention, because of rotation. We recall here that a physical
frame is said to be \textit{time-orthogonal} when there exists at
least one adapted chart in which $g_{0i}=0$: this means that the
lines $x^0=variable$  are orthogonal to the 3-manifold
$x^0=const$. This is an intrinsic property of the physical frame,
i.e. it does not depend on the coordinates used to parameterize
the frame. A way to characterize this feature is the introduction
of  the "spatial vortex tensor", which is a tensor for coordinate
transformation "internal" to the physical frame. The spatial
vortex tensor of the rotating frame $K_{rot}$ is not null: hence
this is not a time-orthogonal frame\footnote{For the definitions
of the "spatial vortex tensor" and of the "internal" coordinate
transformations, together with their applications to the rotating
disk, see \cite{rizzi02}.}. In particular, a global clocks
synchronization is impossible in these non time-orthogonal
frames\cite{landau}.

\subsection{The local spatial geometry of the rotating frame}

Using the metric (\ref{born}) the line element, in the chart
$\left\{ x'^{\mu \ }\right\} $, is written in the general
axis-symmetric form:
\begin{equation}
ds^2=g'_{00}c^2dt'^2+g'_{rr}dr'^2+g'_{\vartheta\vartheta}d\vartheta'^2+g'_{zz}dz'^2+
2g'_{t \vartheta}cdt' d\vartheta' \label{eq:bornline}
\end{equation}

We can introduce the local spatial geometry of the disk, which
defines the proper spatial line element, on the basis of the local
optical geometry. To this end we can use the radar
method\cite{gron3}, \cite{landau}.

\begin{figure}[here]
\begin{center}
\includegraphics[width=7cm,height=7cm]{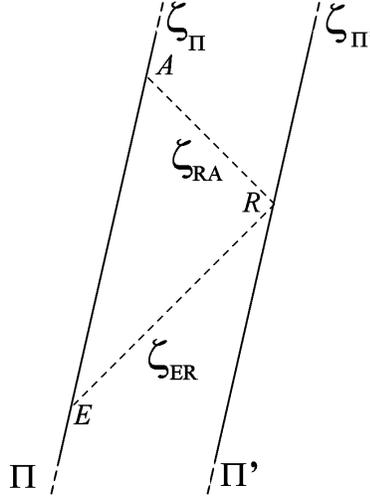}
\caption{\small $\zeta_{\Pi}$ and $\zeta_{\Pi'}$ are the
world-lines of the two neighbouring points $\Pi$ and $\Pi'$ in the
rotating frame, between which a light signal propagates. The event
$E$ corresponds to the emission of the light signal (at time
$t'_R+dt'_E$) which propagates along the null world-line
$\zeta_{ER}$. The events $R$ corresponds to the reflection of the
signal (at time $t'_R$) which, after propagating along the null
world line $\zeta_{RA}$, is absorbed at the space-time event $A$
(at time $t'_R+dt'_A$).} \label{fig:radar}
\end{center}
\end{figure}

Let $\Pi$ be a point in the rotating frame, where a light source,
a light absorber and a clock are lodged; let $\Pi'$ be a near
point where a reflector is lodged. The world-lines of these points
are the time like helices $\zeta_\Pi$ and $\zeta_{\Pi'}$ (see
figure \ref{fig:radar}). A light signal is emitted by the source
in $\Pi$ and propagates along the null world-line $\zeta_{ER}$
toward $\Pi'$; here it is reflected back to $\Pi$ (along the null
world-line $\zeta_{RA}$), where it is finally absorbed. Let
$d\tau$ be the proper time, read by a clock in $\Pi$, between the
emission and absorption events: then, according to the radar
method, the proper distance between $\Pi$ and $\Pi'$ is defined by
\begin{equation}
d \sigma =\frac{1}{2}c d \tau \label{eq:spacegeo1}
\end{equation}

Now, we are going to parameterize these events, using the
coordinates adapted to the rotating frame, in order to obtain the
explicit expression of the proper spatial line element. Let $x'^i$
and $x'^i+dx'^i$ be, respectively, the spatial coordinates of
$\Pi$ and $\Pi'$. The space-time intervals between the events of
emission $E$ and reflection $R$, and between the events of
reflection $R$ and absorption $A$, are null. Hence, by setting
$ds^2=0$ in (\ref{eq:bornline}), we can solve for $dt'$, and
obtain the two solutions:
\begin{eqnarray}
dt'_{E}   & = & \frac{1}{cg'_{00}}
\left(-g'_{t\vartheta}-\sqrt{(g'_{t\vartheta}d\vartheta')^2-g'_{00}
\left(g'_{rr}dr'^2+g'_{\vartheta\vartheta}d\vartheta'^2+g'_{zz}dz'^2
\right) } \right) \label{eq:dtE} \\
dt'_{A}   & = & \frac{1}{cg'_{00}}
\left(-g'_{t\vartheta}+\sqrt{(g'_{t\vartheta}d\vartheta')^2-g'_{00}
\left(g'_{rr}dr'^2+g'_{\vartheta\vartheta}d\vartheta'^2+g'_{zz}dz'^2
\right) } \right) \label{eq:dtA}
\end{eqnarray}
which correspond to the propagation along the two directions
between $\Pi$ and $\Pi'$.\footnote{That is along the null
world-lines $\zeta_{ER}$ and $\zeta_{RA}$.} So, if $t'_{R}$ is the
coordinate time of the reflection event, the coordinate times of
the  emission and absorption events are, respectively,
$t'_{R}+dt'_{E}$ and  $t'_{R}+dt'_{A}$. Consequently, the
coordinate time elapsed between these two events turns out to be
\begin{eqnarray}
\delta t' & \doteq & (t'_{R}+dt'_{A})-(t'_{R}+dt'_{E}) \nonumber
\\
 & = & dt'_{A}-dt'_{E} \nonumber \\
 & = & \frac{2}{cg'_{00}}\sqrt{(g'_{t\vartheta}d\vartheta')^2-g'_{00}
\left(g'_{rr}dr'^2+g'_{\vartheta\vartheta}d\vartheta'^2+g'_{zz}dz'^2
\right) } \label{eq:deltati}
\end{eqnarray}
The corresponding proper time difference is
\begin{equation}
d\tau=\sqrt{g'_{00}}\delta t' \label{eq:dtaudtprimo}
\end{equation}
Hence, using (\ref{eq:dtaudtprimo}), (\ref{eq:deltati}) and the
definition (\ref{eq:spacegeo1}) of the radar spatial proper
distance, we obtain
\begin{equation}
d\sigma=\frac{1}{\sqrt{g'_{00}}}\sqrt{(g'_{t\vartheta}d\vartheta')^2-g'_{00}
\left(g'_{rr}dr'^2+g'_{\vartheta\vartheta}d\vartheta'^2+g'_{zz}dz'^2
\right) } \label{eq:deltasigma}
\end{equation}

Explicitly, by inserting the elements of the metric tensor
(\ref{born}), the proper spatial line element is written as
\begin{equation}
d\sigma^2= dr'^2+\gamma^2 r'^2 d\vartheta'^2+d z'^2
\label{eq:spaceline}
\end{equation}
where $\gamma=\gamma(\omega,r)\doteq 1/\sqrt{1-\frac{\omega^2
r^2}{c^2}}$.

In general this method leads to defining the  proper spatial line
element in the form:
\begin{equation}
d\sigma^2 \doteq \gamma'_{ij}dx'^i dx'^j \label{eq:dl2}
\end{equation}
where  $\gamma'_{ij}$ can be thought of as a "spatial metric
tensor". According to eq. (\ref{eq:spaceline}) the components of
this tensor are
\begin{equation}
\gamma' _{ij}=\left(
\begin{array}{ccc}
1 & 0 & 0 \\
0 & \gamma ^{2}{r'}^{2} & 0 \\
0 & 0 & 1
\end{array}
\right) \mathrm{\ \quad .}  \label{eq:metricagamma}
\end{equation}
If we concern with a disk, where the $z$ coordinate is constant,
we get
\begin{equation}
d\sigma^2=dr'^2+\gamma^2 r'^2 d\vartheta'^2 \label{eq:gammaspace1}
\end{equation}

To complete the description of the rotating frame, it is easy to
check that this frame is rigid according to Born's definition of
rigidity:  a body  moves rigidly if the spatial distance
$d\sigma~=~\sqrt{\gamma'_{ij}dx'^i dx'^j}$ between neighbouring
points of the body, as measured in their successive (locally
inertial) rest frames, is constant in time. This is the case of
the rotating frame $K_{rot}$, since the spatial metric tensor
$\gamma'_{ij}$ does not depend on time.

\section{The "relative space" of a rotating disk} \label{sec:relspace}

\normalsize

In the previous section we have outlined the local spatial
geometry of a rotating disk on the bases of the local optical
geometry\footnote{It is worthwhile to  stress that the use of the
\textit{optical congruence} is meaningful only in any
\textit{local} Minkowskian (tangent) frame, whose space geometry
is actually the geometry of an optical space.}. Namely, when light
rays are used, locally, as standard rods, the line element which
allows measurements of lengths is given by (\ref{eq:spaceline}).
This local spatial geometry is defined at each point of the
rotating frame. However, in order to have the possibility of
confronting measurements performed at different points in the
frame, a procedure to extend all over the disk the local spatial
geometry is required. But this cannot be done in a straightforward
way, because  a rotating frame is not time-orthogonal and hence it
is not possible to choose an adapted charted in which, globally,
the lines $x^0=var$ are orthogonal to the 3-manifold $x^0=const$
(each of which is described by the metric
(\ref{eq:metricagamma})). In other words, a global foliation,
which would lead "naturally" to the definition of the space of the
disk, is not allowed.

Nevertheless, if we shift the context, from the ill defined notion
of space of the disk thought of  as a subspace or a submanifold
embedded in space-time, to a definition which has a well defined
and operational meaning, we are lead to the concept of the
"relative space". To this end, first of all let us start from a
more formal description of the local splitting of space-time that
allows us to write the local spatial metric
(\ref{eq:metricagamma}). Let $\zeta$ be a time-like helix,
describing the evolution of a point of the disk, and let $P$ be an
event which belongs to $\zeta$. We can identify a 1-dimensional
vector space $\Theta_P$ ("time direction"), which is spanned by
the vector tangent to $\zeta$ in $P$, and a 3-dimensional vector
space $\Sigma_P$ ("space platform"), normal to $\Theta_P$;
$\Sigma_P$ is endowed with the metric (\ref{eq:metricagamma}).
Hence the space tangent to $P$ has a splitting in the form
$T_{P}=\Theta _{P}\oplus \Sigma _{P}$.

This local splitting has an important property: \textbf{\
}\textit{the splitting }$T_{P}=\Theta _{P}\oplus \Sigma
_{P}$\textit{\ and the spatial metric tensor }$\gamma'_{ij}(P)$
are invariant along the lines of congruence $\Gamma$, i.e. they
are dragged along the world evolution. This property is strictly
related to the rigidity of the rotating frame, namely it depends
on the fact that the metric (\ref{born}) is globally stationary
and the metric (\ref{eq:metricagamma}) is locally static. In
particular, it can be explained in terms of isometries, i.e.
Killing fields in the submanifold $r=const$ of the space-time(see
\cite{rizzi02}). As a consequence it is  possible to define a
one-parameter group of diffeomorphisms with respect to which both
the splitting $T_{P}=\Theta _{P}\oplus \Sigma _{P}$ and the space
metric tensor $\gamma'_{ij}(P)$ are invariant. The lines of
$\Gamma$ constitute the trajectories of this "space~$\oplus$~time
isometry".\\

This fact suggests a procedure to define an \textit{extended
3-space}, which we shall call \textit{relative space}. Naively, it
can be thought of as the union of the infinitesimal (local) space
platforms, but a more rigorous
definition can be given:\\

\textbf{Definition. } Each element of the relative space is an
equivalence class of \textit{points and of space platforms}, which
verify this equivalence relation:\newline

RE\label{relequiv}:\textit{\ \noindent `` Two points (two space
platforms) are equivalent if they belong to the same line of the
congruence ''}.\newline

That is, the \textit{relative space} is the "quotient space" of
the world tube of the disk, with respect to the equivalence
relation RE, among points and space platforms belonging to the
lines of the congruence $\Gamma$.

This definition simply means that the relative space is the
manifold whose ''points'' \textit{are} the lines of the
congruence. Our definition emphasizes the role of the space
platforms: the reference frame defined (as above) by the relative
space coincides \textit{everywhere} with the local rest frame of
the rotating disk.

In other words, the relative space is a formal tool which allows a
connection among all the local optical geometries that are defined
in the neighbourhood of each point of the space.  As a
consequence, space  measurements globally defined in the relative
space reduce, immediately, to standard measurements in any local
frame, co-moving with the disk. In this way, a natural procedure
to make a comparison between observations performed by observers
at different points is available. The physical context in which
these distant observations are made, is defined, both from a
mathematical and operational point of view, by the relative space.

We want to stress, again, that it is not possible to describe the
relative space in terms of space-time foliation, i.e. in the form
$x^{0}=const$, where $x^{0}$ is an appropriate coordinate time,
because the space of the disk, as we saw before, is not
time-orthogonal. Hence, thinking of the space of the disk as a
sub-manifold or a subspace embedded in the space-time is
misleading: instead, the space of the disk, defined by the
relative space, must be thought of as a quotient space. If we long
for some kind of visualization, we can think of the relative space
as the union of the infinitesimal space platforms, each of which
is associated, by means of the request of M-orthogonality, to one
and only one of lines of the congruence.

\section{Measurements  in the relative space} \label{sec:measure}

After having introduced  the relative space and described its
geometry, we come back  to our original purposes, i.e. the
solution of the Ehrenfest's paradox.

Space measurements are locally performed by a rotating observer by
means of the spatial metric  tensor $\gamma^{\prime}_{ij}$:

\begin{equation}
\gamma _{ij}^{\prime }=\left(
\begin{array}{ccc}
1 & 0 & 0 \\
0 & \gamma ^{2}{r'}^{2} & 0 \\
0 & 0 & 1
\end{array}
\right)   \label{eq:metricagamma1}
\end{equation}

As a consequence, the length of an infinitesimal segment on the
rim of the circumference is
\begin{equation}
dl^{\prime}_{\Sigma}=\left(\sqrt{\gamma_{ij}'(\omega,r')dx'^i
dx'^j} \right)_{r'=R,z'=cost}=\gamma(\omega,R)R d\vartheta'
\label{eq:dlsigma}
\end{equation}
From (\ref{catt})$_{III}$ it follows that at fixed coordinate time
of the inertial frame $K$, $d\vartheta'=d\vartheta$. Consequently,
the angle all around the periphery of the disk,
measured on it, is equal to $2\pi$: $\vartheta' \in [0,2\pi]$.\\
Hence, the measurement of the circumference on the rim of the
disk, performed by the rotating observer, turns out to be :
\begin{equation}
l^{\prime}_{\Sigma}= 2 \pi R \gamma  \label{eq:lsigma}
\end{equation}
We want to stress  that the dilation appearing in
(\ref{eq:lsigma}) has a pure kinematical origin and by no means it
ought to be interpreted as a dynamical process, involving the
structure of the disk, as some authors claimed in the
past\cite{cavalleri}. The meaning of this dilation will be made
clear in next section.

For the observer in the inertial frame the length given in eq.
(\ref{eq:dlsigma}) appears contracted by the standard factor
$\gamma^{-1}$:
\begin{equation}
dl=\gamma^{-1}dl^{\prime}_{\Sigma}  \label{eq:dlI}
\end{equation}
Since
\begin{equation}
dl=\gamma^{-1}\gamma R d\vartheta^{\prime}= Rd\vartheta^{\prime}
\label{eq:dlI2}
\end{equation}
we obtain that, in correspondence to the measure of the
circumference given in eq. (\ref{eq:lsigma}), performed by the
rotating observer, the inertial observer measures a length $2\pi
R$, as expected, since the space of the inertial frame $K$ is
Euclidean.

The expression (\ref{eq:lsigma})  is in agreement with the fact
that the space geometry of the disk is not Euclidean: a curvature
tensor can be computed and it is not null\footnote{It is
interesting to notice that this confirms Einstein's early
intuition\cite{einstein}: he suggested that rotation must distort
the Euclidean geometry of the platform, so that the geometry of
the inertial frame remains Euclidean. Indeed the rotating disk was
just an euristic tool in order to investigate the possibility that
the geometry of the Minkowskian space-time could be distorted
by a gravitational field\cite{stachel}.}.\\

\textbf{Remark 1.}  In the past, different authors, like
Berenda\cite{berenda}, Arzeli\`{e}s\cite{arzelies}, Gr\o
n\cite{gron1} calculated the curvature of the space of the disk.
It is worthwhile to notice that  they did not define the proper
geometrical context in which their calculation were performed;
moreover, their calculations   do not rely upon the use of a
splitting technique: they just computed the components of the
curvature tensor of the \textit{space-time} which have all spatial
indices ($R_{ijkl}$), and they referred to it as the
\textit{space} curvature tensor. On the other hands, we showed
 that the relative space is the natural
context in which space measurements are performed and  we computed
its curvature (see \cite{rizzi02}) using the Cattaneo's splitting
technique. Nevertheless their results are equal to ours, and this
is due to the fact that, for the rotating disk in uniform motion,
the physical frame  is stationary; however, things are different
for those physical reference frames which lack symmetries, such as
the axis-symmetry and the
stationarity of the rotating disk.\\

\section{Lengths in the relative space}\label{sec:lengths}

We want to clarify, here, the origin of the dilation appearing in
(\ref{eq:lsigma}), which, as we showed, is ultimately responsible
for the solution of   the Ehrenfest's paradox. In order to do
that, it is useful to analyze what happens to "standard lengths"
when they undergo an acceleration process.\\

\begin{figure}[top]
\begin{center}
\includegraphics[width=8cm,height=8cm]{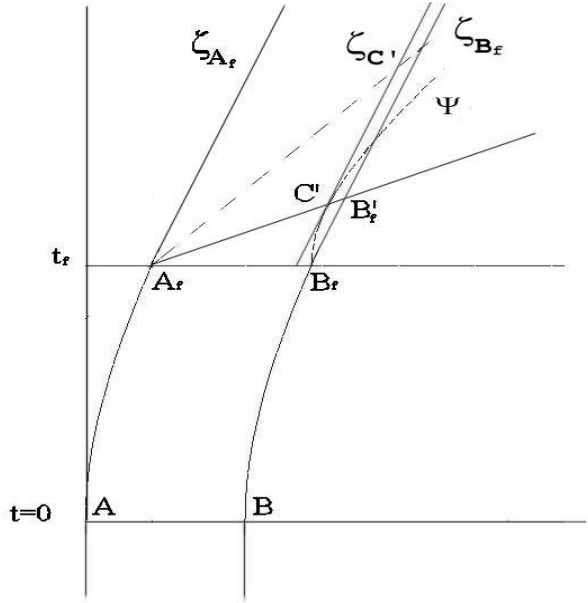}
\caption{\small The world-strip $(\zeta_{A_f},\zeta_{C'})$ of a
length $\lambda_0=AB=A_f B_f$, at rest on the rim, does not cover
entirely the world-strip $(\zeta_{A_f},\zeta_{B_f})$ of this
length, as measured in the inertial frame.} \label{fig:figura}
\end{center}
\end{figure}

Let us consider the world-strip of an infinitesimal piece of the
rim of the disk, which is at rest until $t=0$ in the inertial
frame $K$ (see figure \ref{fig:figura}).

When $t=0$, the disk starts being accelerated in such a way that
all points of its rim have identical motion, as observed in $K$.
If $I=[0,t_f]$ is the interval  representing the period of time
during which acceleration acts, $\forall t \in I$ the acceleration
distribution of all points of the rim is the same, as observed in
the inertial frame $K$. From a pictorial point of view, this means
that the world lines of all points of the rim are congruent (i.e.
superimposable). During the acceleration period, the disk is not
Born-rigid although it appears rigid in $K$. This means that,
depending on the simultaneity criterium in the inertial frame, the
length of the infinitesimal piece of the rim is always congruent
with the starting segment $AB$; in particular, when $t=t_f$, it is
represented by the segment $A_f B_f$. On the other hands, from the
point of view of the local observer at rest on the rim, whose
world line passes through $A$ when $t=0$, the simultaneity
criterium is not defined by the family of straight lines parallel
to $AB$, but it varies at each instant, depending on the velocity
(in $K$) of the rim itself. Namely, when the acceleration period
finishes, the piece of the rim  is represented by the segment $A_f
B'_f$, in the local co-moving frame associated with $A_f$. Let us
put $A_f B_f=AB=\lambda_0$, where $\lambda_0$ is the wavelength of
a monochromatic radiation emitted by a source at rest in
$K$\footnote{Since $AB$ is infinitesimal, the monochromatic
radiation must be chosen in such a way that $\lambda_0$ is very
small when compared with the length of the circumference.}. The
$M$-circumference of radius $\lambda_0$, with center in $A_f$,
whose equation is
\begin{equation}
\Psi \equiv \{P\in M^{2}:A_{f}P=\lambda _{0}\}
\label{eq:Mcircumference}
\end{equation}
can be built by  considering, in each reference frame, the
wavelength of the given radiation, emitted by a source at rest in
that frame. This $M$-circumference, which is a hyperbola in the
Minkowskian plane, intersects the segment $A_{f}B_{f}^{\prime }$
in $C^{\prime }$, and we obtain
\begin{equation}
A_{f}B_{f}^{\prime }=A_{f}C^{\prime }\gamma =\lambda _{0}\gamma
>\lambda _{0} \label{eq:Mcirc}
\end{equation}
This relation means that the world-strip
$(\zeta_{A_f},\zeta_{C'})$ of a length $\lambda_0$, at rest on the
rim, does not cover entirely the world-strip
$(\zeta_{A_f},\zeta_{B_f})$ of this length, as measured in $K$
(see figure \ref{fig:figura}). From a physical point of view,
equation (\ref{eq:Mcirc}) shows that each element of the periphery
of the disk, of proper length $\lambda_0$, is stretched during the
acceleration period. This a purely kinematical result of our
acceleration program.

However, this result remains valid if one takes into account the
interactions among the physical points of the rim (f.i. those
interactions which ensure rigidity in the phase of stationary
motion). In particular, during and after the acceleration period,
each point of the disk is subject to both radial and tangential
stresses; the former maintain each point on the circumference
$r=r_0$, while the latter ought to give zero resultant, because of
the axial symmetry: \textit{each point is pulled in the same way
by its near points, in both directions}. As a consequence, the
elongation of every element of the rim, due to tidal forces
experienced during the acceleration period, remains even when
acceleration finishes\footnote{Let us point out that a shortening
of the elements of the rim, due to Hooke's law, cannot be invoked,
since these elements have not free
endpoints.}.\\

\normalsize

\textbf{Remark } The arguments given treat the disk as a set of
non interacting particles. The only constraint is that every
particle must move along a circular trajectory, with a given law
of motion, according to a kinematical definition of rigidity in
$K$ .\\

From the considerations above, it follows that the dilation which
is responsible for the solution of the Ehrenfest's paradox  has a
pure kinematical origin. The enlargement of the rod (assumed as a
standard rod), in the rest frame  at the end of acceleration
phase, is due to the change of the simultaneity criterium. In
figure \ref{fig:figura} this is represented by the change in the
slope of the infinitesimal space platforms which are associated,
by means of the request of $M$-orthogonality, to the lines of the
congruence $\Gamma$.

\section{Conclusions} \label{sec:conclusion}

We introduced the concept of relative space in order to make clear
the fundamental processes of measurements that take place on a
rotating platform, or more generally, in a rotating frame. In
particular, space measurements are involved in the solution of the
Ehrenfest's paradox. Often, in the literature on these subjects,
misunderstandings arise, and we believe that they rely on the lack
of clear and self consistent definitions of the fundamental
concepts used.

According to us, the concept of   relative space makes clear, in a
mathematical way, the physical context in which measurements are
performed by rotating observers. Even though a global
\textit{isotropic} $1+3$ splitting of the space-time is not
allowed when dealing with rotating observers, space measurements
are defined globally in the relative space, and reduce in a
straightforward way to the standard  measurements in any local
frame, co-moving with the disk.

In the relative space we have been able to outline the solution of
the Ehrenfest's paradox, evidencing its kinematical origin, which
is related to the way of measuring lengths in a rotating platform.
Furthermore, the geometry of the space of the disk, which follows
from our assumptions, turns out to be non Euclidean, according to
Einstein's early intuition. The solution of the  Ehrenfest's
paradox, that we outlined in this paper, is strongly dependent
both on a proper definition of the physical space of the disk,
i.e. the relative space, and on a proper choice of the congruence
adopted to perform the measurements in such a space, i.e. the
(local)  optical congruence.

In conclusion, it appears  that  SRT, even when applied to
rotating platforms, is self-consistent and does not raise
paradoxes, provided that proper definitions of geometrical and
kinematical entities are adopted.

We believe that our approach to the study of these apparently
paradoxical problems leads to a deeper understandings of the very
foundations of the theory and evidences, in a clear way, some
operational
aspects of the measurement processes involved.\\

\textbf{Acknowledgements } The author is grateful to Professor
Guido Rizzi, for valuable suggestions and comments.

\pagebreak

\end{document}